\documentclass[aps, prb, longbibliography, reprint]{revtex4-1}
\usepackage{graphicx}
\usepackage{color}
\begin{document}
\newcommand{\eqref}[1]{Eq.~(\ref{eq:#1})}
\newcommand{\figref}[1]{Fig.~\ref{fig:#1}}
\newcommand{\secref}[1]{Sec.~\ref{sec:#1}}
\newcommand{\deffig}[4]{%
\begin{figure}[htb]
\includegraphics[width=#2\textwidth]{#3}
\caption{#4}
\label{fig:#1}
\end{figure}
}
\title{
Entanglement branching operator
}
\author{Kenji Harada}
\affiliation{Graduate School of Informatics, Kyoto University, Kyoto 606-8501, Japan}
\begin{abstract}
  We introduce an entanglement branching operator to split a composite
  entanglement flow in a tensor network which is a promising
  theoretical tool for many-body systems. We can optimize an
  entanglement branching operator by solving a minimization problem
  based on squeezing operators. The entanglement branching is a new
  useful operation to manipulate a tensor network. For example,
  finding a particular entanglement structure by an entanglement
  branching operator, we can improve a higher-order tensor
  renormalization group method to catch a proper renormalization flow
  in a tensor network space. This new method yields a new type of
  tensor network states. The second example is a many-body
  decomposition of a tensor by using an entanglement branching
  operator. We can use it for a perfect disentangling among
  tensors. Applying a many-body decomposition recursively, we
  conceptually derive projected entangled pair states from quantum
  states that satisfy the area law of entanglement entropy.
\end{abstract}
\maketitle
%
%
\section{Introduction}
In the last decade, a tensor network grows a new promising theoretical
tool for treating many-body systems. A novel property of a quantum
state like a topological order\cite{Schuch:2010jp} and a symmetry
protected topological order\cite{Chen:2011vg} can be explicitly
constructed by tensor networks. Tensor networks help us to understand
novel properties of a quantum state as a specific property of a
tensor.  Based on the area law of entanglement entropy, we can define
a general class of quantum states as a tensor network which has a
special structure. For example, projected entangled pair states
(PEPS)\cite{Verstraete:2006jk} and multi-scale entanglement
renormalization ansatz (MERA)\cite{Vidal:2007kx}. We can control the
quality of these tensor network states through the degrees of freedom
on tensor indexes. Thus, we can use a tensor network as a promising
variational wave function for strongly correlated materials. We can
also define tensor network formulation of many-body problems. It gives
us a new perspective way to treat many-body problems. For example,
contracting a tensor network with controllable accuracy, we can
systematically calculate the property of many-body systems.

To optimize a tensor in a tensor network and to calculate a
contraction of a tensor network, novel numerical algorithms for a
tensor network have been proposed in the last
decades\cite{Nishino:1996bw, Vidal:2007hx, Levin:2007ju, Xie:2012iy,
Evenbly:2015cs, Yang:2017hj}. They help us to understand the
properties of strongly correlated materials numerically (for example,
see Refs.\cite{Mila:2014eu, Corboz:2014ba, Evenbly:2010hh,
Harada:2012ht}). Thus, the development of tensor algorithms is highly
active. However, the types of operations in a tensor network algorithm
are limited.

In this paper, we will propose a new tensor operation which is called
an entanglement branching (EB). The EB is to split a composite
entanglement flow in a link of a tensor network. We will explicitly
introduce an EB operator in a tensor network.

In \secref{TN}, we will briefly introduce tensor networks, tensor
operations, and tensor network algorithms. In \secref{EB}, we will
define an EB operator and a local problem to optimize it. In
\secref{APP}, we will show two applications of the EB operation. One
is an improvement of the higher-order tensor renormalization group
(HOTRG)\cite{Xie:2012iy} to catch a proper renormalization flow in a
tensor network space. The other is a many-body decomposition of a
tensor. In \secref{SUM}, we will conclude and discuss our results.

\section{Tensor networks, tensor operations, and tensor network
algorithms}
\label{sec:TN}
A tensor network is a theoretical tool to describe correlations
between elements in a system. At first, we will introduce a useful
graphical notation for tensor networks. Secondly, we will introduce
conventional operations in tensor networks. Finally, we will show an
example of tensor network algorithms.

\deffig{TN}{0.48}{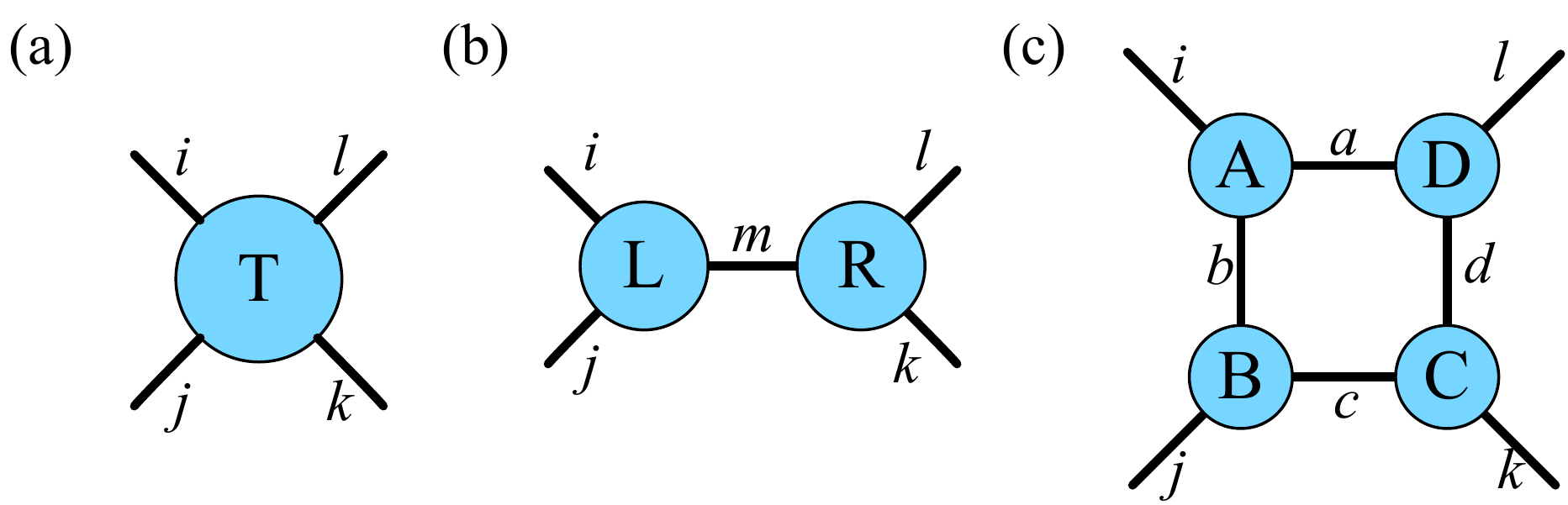}{(a) A graphical representation of a tensor
  $T$. Four lines represent tensor indexes $i$, $j$, $k$, and $l$.
  (b) A graphical representation of a tensor contraction between $L$
  and $R$.  A link between two tensors denotes a pair of contracted
  indexes. For example, a link $m$ represents a tensor contraction for
  the tensor $L$ index and the tensor $R$ index. Thus, this diagram
  represents a composite tensor $(LR)$ as
  $(LR)_{ijkl} = \sum_m L_{ijm}R_{mkl}$.  Applying a matrix
  decomposition, we can decompose a tensor into two tensors with a
  tensor contraction as shown in this diagram.  (c) A tensor network
  which consists of four tensors, $A$, $B$, $C$, and $D$. We call it a
  four-body tensor network.}

\figref{TN} shows a graphical representation of a tensor and a tensor
network. The object in \figref{TN}(a) represents a tensor $T$. Each
line from the object represents each index of $T$. The link (labeled
$m$) between tensor $L$ and $R$ in \figref{TN} (b) represents a tensor
contraction for a tensor $L$ index and a tensor $R$ index.  Thus, the
whole of \figref{TN} (b) represents a composite tensor $(LR)$ :
$(LR)_{ijkl} \equiv \sum_m L_{ijm} R_{mkl}$. \figref{TN} (c)
represents a complex composite tensor which consists of four
tensors. Since these diagrams visually seem to be networks of tensors,
they are called tensor networks.

A quantum state is defined in a tensor product space of localized
Hilbert spaces. Thus, if we can regard a tensor index as the degrees
of freedom in a localized Hilbert space, the wave function is written
as a tensor. For example, we can regard four indexes $i$, $j$, $k$,
and $l$ in \figref{TN} (a) (b) (c) as the physical degrees of freedom
in a four-body system. A quantum state defined by a tensor network is
called a tensor network state. We can use a tensor network state to
represent a novel quantum state explicitly\cite{Schuch:2010jp,
Chen:2011vg}. If a tensor network state satisfies the area law of
entanglement entropy as like PEPS and MERA, we can use it as a
variational wave function which is systematically controllable. In
general, an entanglement flows through a link of a tensor network. If
we consider a cut of a tensor network to decompose physical indexes
into two groups, the entanglement entropy of the decomposed sub-system
is less than $\sum_{i \in \mbox{cut}} \log(D_i)$, where $D_i$ is the
degrees of a link $i$ in a cut. Thus, a link $i$ maximally contributes
$\log(D_i)$ to an entanglement entropy. The minimum cut defines a
limit of an entanglement entropy of a tensor network state. Therefore,
the property of an entanglement entropy in a tensor network state
depends on the geometrical structure of a tensor network.

There are two basic operations to manipulate a tensor network. One is
a tensor contraction, and the other is a tensor decomposition. We
calculate a tensor contraction in a tensor network to obtain a new
composite tensor. For example, from \figref{TN}(b) or (c) to (a).
Currently, the tensor decomposition is simply based on a matrix
decomposition. However, the matrix-based tensor decomposition has a
limit of a transformation of tensor network topology. For example,
using a matrix-based tensor decomposition, we can transform a tensor
$T$ in \figref{TN}(a) to a tensor network of $L$ and $R$ in
\figref{TN}(b). However, we cannot transform \figref{TN}(a) to
(c). The matrix-based tensor decomposition produces only a two-body
tensor network.  Even the higher-order singular value decomposition
(HOSVD) has the same limit that can be regarded as the sequence of
two-body decomposition. The EB operation proposed in this paper will
resolve this limit (see \secref{MBD}).

Various types of tensor network algorithms have been proposed in the
last decades\cite{Nishino:1996bw, Vidal:2007hx, Levin:2007ju,
Xie:2012iy, Evenbly:2015cs, Yang:2017hj}. Here, we will give a brief
introduction of the HOTRG algorithm\cite{Xie:2012iy} as a typical
tensor network algorithm. A partition function of a classical or
quantum system can be written by a grid-type tensor network as shown
in \figref{HOTRG}(a). HOTRG algorithm approximately makes a
coarse-grained tensor by inserting projection operators as shown in
\figref{HOTRG}(b). We calculate projection operators from a HOSVD of a
tensor $T$. Calculating tensor contractions among two $T$s and two
$P$s, we obtain a coarse-grained tensor $T'$ in \figref{HOTRG}(c). The
number of tensors in the new tensor network is half. Thus, HOTRG
algorithm is a real-space renormalization on a tensor
network. Repeating this procedure with changing a direction, we
finally obtain a single tensor. A trace of a coarse-grained tensor is
an approximation of all tensor contractions in the original tensor
network. In general, as like the HOTRG algorithm, a procedure in a
tensor network algorithm is a combination of tensor contractions and
matrix-based tensor decompositions.
\deffig{HOTRG}{0.48}{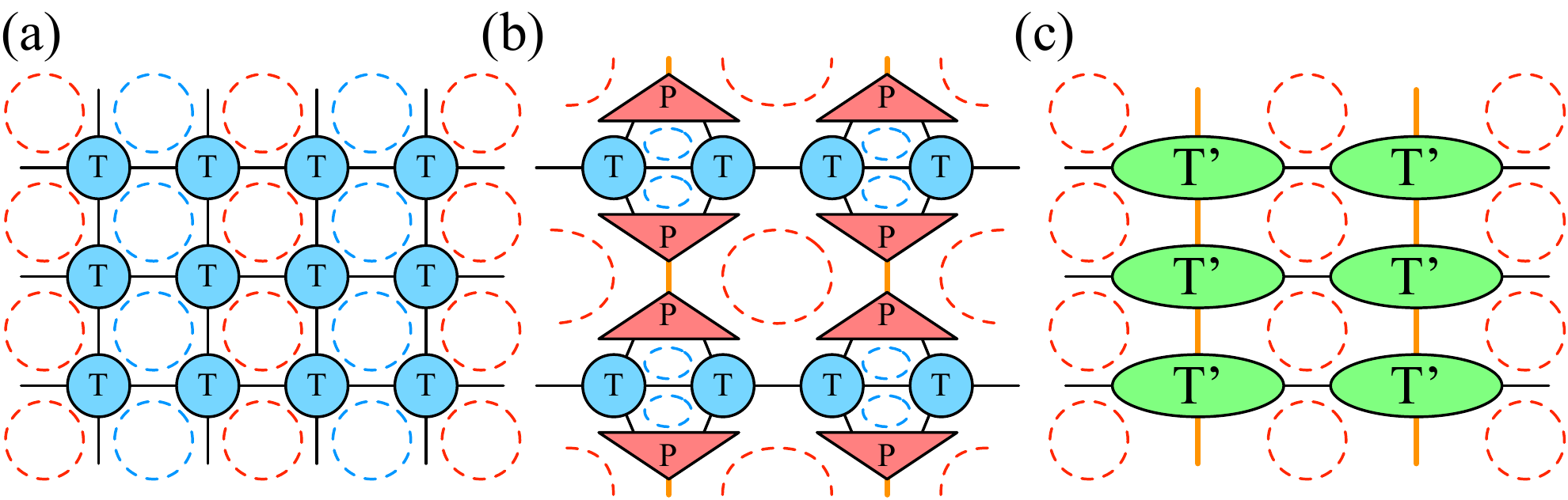}{(a) A tensor network representation of
  a partition function of a two-dimensional classical system or a
  one-dimensional quantum system. Dotted circles denote a short-range
  loop entanglement structure. Red and blue colors denote remained and
  erased entanglement flows by a HOTRG procedure, respectively. (b) A
  coarse-graining procedure in a HOTRG algorithm. A projection
  operator $P$ is usually determined by a HOSVD.  (c) A new
  renormalized tensor network for (a). A new tensor $T'$ is the result
  of a tensor contraction of two $T$s and two $P$s in (b).}

\section{Entanglement branching}
\label{sec:EB}
A link in a tensor network carries an entanglement flow. The
entanglement flow may be composite. For example, the entanglement flow
in a link $m$ of a tensor $L$ in \figref{TN}(b) may include two
entanglement flows from $i$ and $j$. Here, we consider a splitting of
a composite entanglement flow in a link as EB.

\deffig{BR}{0.48}{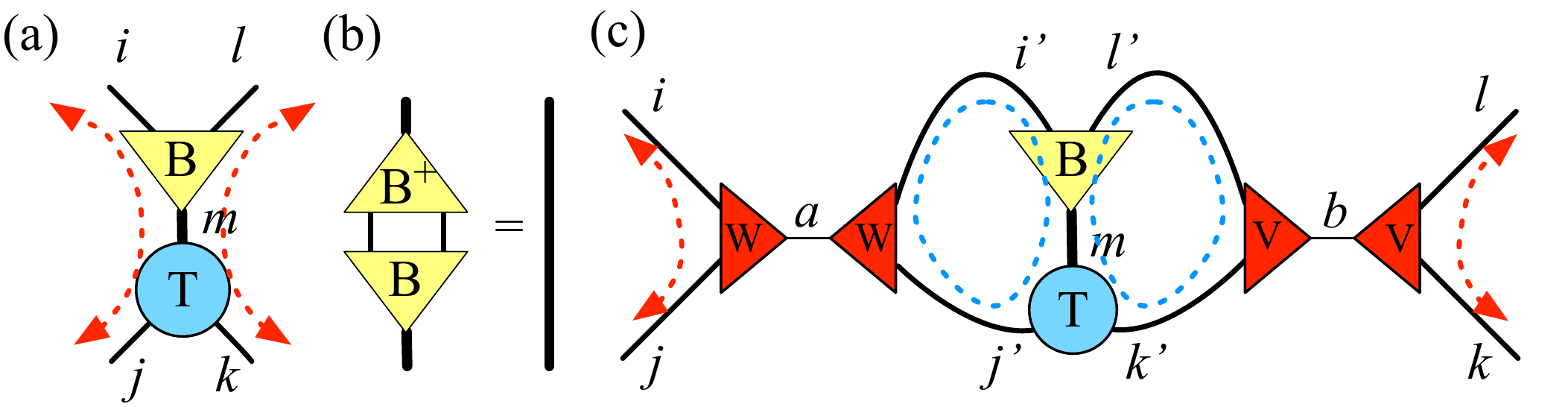}{(a) An EB operator $B$ splits a composite
entanglement flow on a link $m$ into left and right directions. (b) An
isometric property of an EB operator $B$.  (c) A tensor network with
squeezing operators $w$ and $v$.}

To define the EB operation explicitly, we introduce an isometric EB
operator for a link of a tensor. For the sake of simplicity, we will
discuss a splitting of a composite entanglement flow which consists of
two entanglement flows.  \figref{BR}(a) shows an EB operator $B$ which
splits a composite entanglement flow on a link $m$ into upper left and
right links ($i$ and $l$). Here, based on a real space geometry, we
consider that upper left and right links ($i$ and $l$) of $B$ should
carry entanglement flows from lower left and right links $j$ and $k$
of $T$, respectively (see two dotted curves in \figref{BR}(a)).

We can freely insert a pair of EB operator, $B$, and $B^\dagger$, on a
link in a tensor network without approximation, because $B$ is
isometric (see \figref{BR}(b)). The insertion directly redesigns a
tensor network to add new links which carry split entanglement
flows. It gives us a new freedom to transform the topology of a tensor
network as discussed in \secref{MBD}.

To find an appropriate EB operator for a target link of a tensor $T$,
we can use squeezing operators. Here, we consider a new tensor network
in \figref{BR}(c).  Tensors $w$ and $v$ in \figref{BR}(c) are
projection operators.  If an entanglement flow from a lower left link
$j'$ of $T$ passes to an upper left link $i'$ of $B$ in
\figref{BR}(c), we can construct a loop entanglement flow among $T$,
$B$, and $w$ (see a left dotted loop in \figref{BR}(c)).  Thus, we can
squeeze the degrees of freedom on a link $a$ without increasing the
distance between two tensor networks in \figref{BR}(a) and (c). We can
also squeeze that on a link $b$ by the combination of $T$, $B$ and
$v$.  Therefore, if we can optimize a branching tensor $B$ and
projection operators $w$ and $v$ to squeeze the degrees of freedom of
links $a$ and $b$ with minimizing the distance of tensor networks in
\figref{BR}(a) and (c), then the optimized tensor $B$ is an
appropriate EB operator. The minimization of the distance between
tensor networks in \figref{BR}(a) and (c) is a local optimization
problem which depends only on $T$. To optimize tensors $B$, $w$, and
$v$, we can use an iteration method in Appendix \ref{ap:iteration}.

We can extend the definition of an EB operator for a composite
entanglement flow which consists of multiple entanglement flows than
2. The optimization problem can be generalized for such case
straight-forwardly.

\section{Applications of entanglement branching}
\label{sec:APP}
The EB operation is a new freedom to manipulate a tensor network
because it can split a composite entanglement flow on a link in a
tensor network. In this section, we will introduce two applications of
the EB operation.

\subsection{Improvement of HOTRG algorithm}
\label{sec:BHOTRG}
We introduced the HOTRG algorithm in \secref{TN} as an example of
tensor network algorithms. The HOTRG algorithm approximately
calculates all tensor contractions in a grid-type tensor network
(\figref{HOTRG}(a)). We can apply it to calculate the partition
function of classical and quantum many-body systems because a
grid-type tensor network is a tensor network representation of a
partition function.

While we can regard the HOTRG algorithm as a real-space
renormalization group method on a tensor network as shown in
\figref{HOTRG}, it may not be a proper real-space renormalization. In
an ideal real-space renormalization, the effect of entanglements under
a new cut-off scale should be renormalized. Thus, entanglement
structures in a renormalized scale should be disappeared after a
real-space renormalization. However, the HOTRG algorithm cannot erase
a loop entanglement structure in a renormalized scale. Dotted loops in
\figref{HOTRG}(a) mean loop entanglement structures in a tensor
network.  Here, we assume that the entanglement of tensor $T$ has a
corner double-line (CDL) structure. Because loop entanglement
structures are defined in a renormalized scale, they should be
disappeared in a new renormalized tensor network of
\figref{HOTRG}(c). While we can remove half of all loop entanglements
by projection operators $P$ in \figref{HOTRG}(b), half of them remains
in a new renormalized tensor network as shown in
\figref{HOTRG}(c). Therefore, a coarse-grained tensor by the HOTRG
algorithm is not a proper renormalized tensor. There is the same
problem in the tensor renormalization group (TRG) algorithm proposed
by Levin and Nave\cite{Levin:2007ju} which is the first real-space
renormalization group method for a grid-type tensor network. In fact,
the invariant entanglement structure for these algorithms is CDL(see Ref. \cite{Ueda:2014gl}). The
idea to erase entanglements in a renormalized scale was firstly
pointed by Gu and Wen \cite{Gu:2009}. However, their
tensor-entanglement-filtering renormalization algorithm cannot
correctly erase entanglements in a renormalized scale near a critical
point. Evenbly and Vidal\cite{Evenbly:2015cs} proposed the use of
disentangler tensors introduced in MERA to improve the TRG
algorithm. Their tensor network renormalization (TNR) algorithm showed
the expected property of an ideal real-space renormalization even at a
critical point. Finally, the importance to erase entanglements in a
renormalized scale was confirmed. In the following, we will consider
the similar improvement of the HOTRG algorithm by the use of EB.

The HOTRG procedure remains a part of loop entanglement flows which
pass through four tensors around plaquettes (see dotted circles in
\figref{HOTRG}(a)). To catch the entanglement flow, we need to split a
part of a composite entanglement flow on a link which belongs to a
loop entanglement structure. Thus, we introduce an EB operator $B$ on
a link $m$ as shown in \figref{BHOTRG}(a). Since the contraction of
$B$ and $B^\dagger$ is identity, we can freely insert the pair into a
link $m$. The purpose of inserting the EB operator is to catch an
entanglement flow which constructs a loop entanglement structure
through the nearest neighbor tensor (see a dotted curve in
\figref{BHOTRG}(a)). To find an appropriate EB operator, the position
of squeezing operators in an optimization problem is important. Here,
our purpose is to split the dotted entanglement flow in
\figref{BHOTRG}(a). If we insert squeezing operators on a left
horizontal link from a tensor $T$ connected to an EB operator $B$ in
\figref{BHOTRG}(a), all entanglement flows from the left horizontal
link to the link $m$ are split into a link $j$. However, our target is
an entanglement flow only in the shortest scale, not one in all
scales. \figref{BHOTRG}(b) shows an effective position of squeezing
operators to select only the target entanglement flow. The
optimization problem of the EB operator $B$ is a minimization of a
distance between two tensor networks, \figref{BHOTRG}(a) and (b). In
general, an entanglement flow on a link is not perfectly
composite. Even then, suitable squeezing operators in an optimization
problem increase a ratio of a target entanglement component.

If we have an EB operator $B$ to split an entanglement flow which
constructs a loop entanglement structure, we can erase it by the
conventional HOTRG procedure. We firstly gather the target
entanglement flow in a tensor by SVD-decomposing the tensor network in
\figref{BHOTRG}(c) into two tensors $L$ and $R$ in
\figref{BHOTRG}(d). For simplicity, we assume a vertical flip symmetry
of $T$. We set the bond dimension of a link between $L$ and $R$ as
that of a horizontal link between two $T$s. In general, the
SVD-decomposition may cause a truncation error. In the case of CDL
tensors, the target entanglement flow is confined in the tensor $L$ in
\figref{BHOTRG}(d). Between $L$ and $R$, there is no entanglement flow
which constructs the shortest loop entanglement structure. Thus, there
is no truncation in the SVD-decomposition into $L$ and $R$. If we
apply a coarse-graining procedure in the HOTRG algorithm to the
combination of $R$ and $L$ as shown in \figref{BHOTRG}(e), we can
erase two loop entanglement structures by a single projection operator
$P$. Finally, there is no loop entanglement structure in a new tensor
network of \figref{BHOTRG}(f). In summary, using an EB operator $B$,
we define new tensors $L$ and $R$ from two tensors $T$. Applying the
HOTRG algorithm to new tensor $R$ and $L$, we can erase all loop
entanglement structures in a renormalized scale.  Therefore, this
procedure may catch a proper renormalized flow in a tensor network
space.

\deffig{BHOTRG}{0.48}{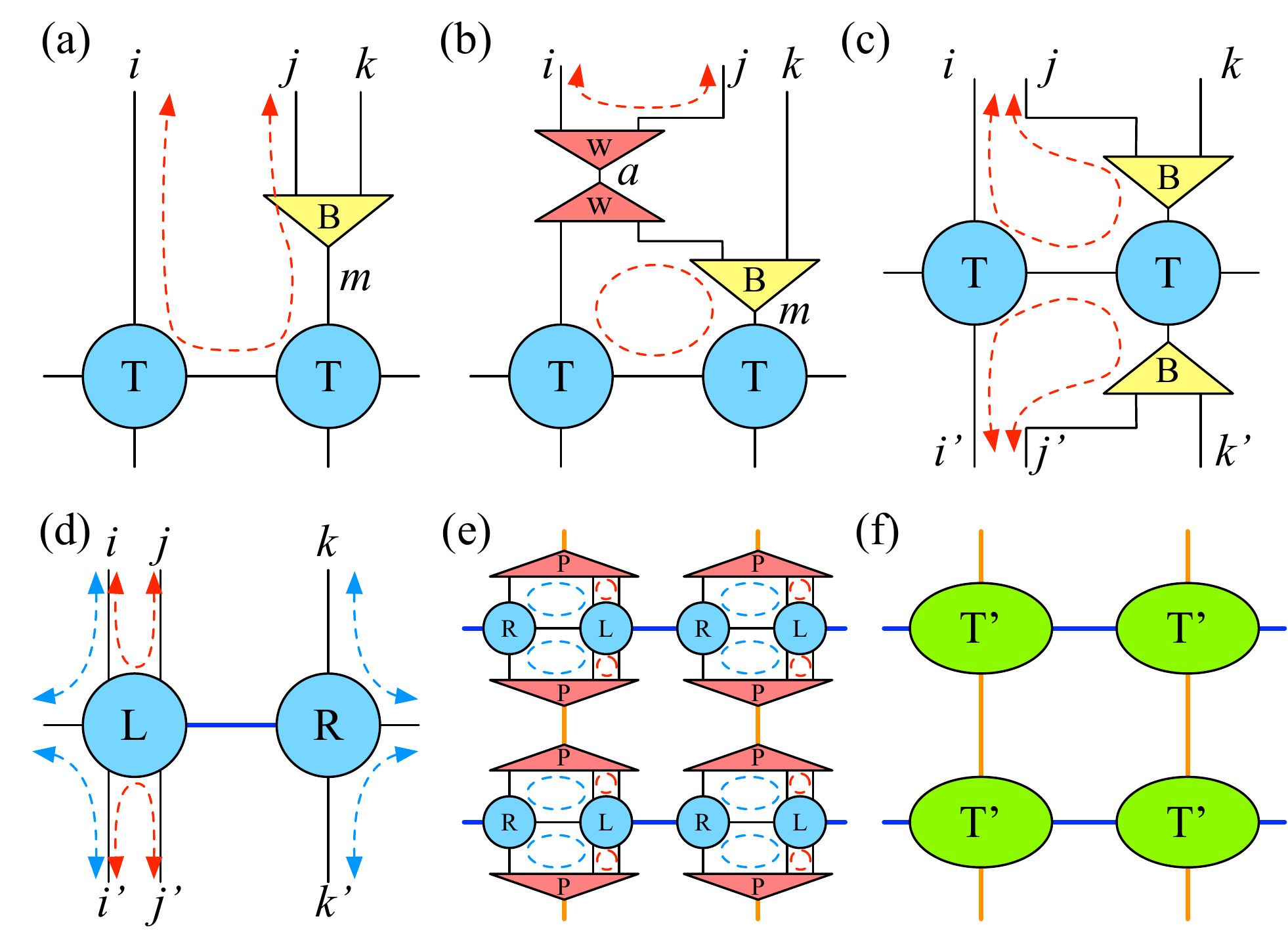}{The HOTRG algorithm with EB
  operations. Red and blue colors denote remained and erased
  entanglement flows by the original HOTRG procedure, respectively.
  (a) An EB operator $B$ to separate a short-range entanglement flow.
  (b) A tensor network with squeezing operators $w$ for the
  optimization of $B$.  (c) Insertion of EB operators $B$ in a
  grid-type tensor network.  (d) New tensors $L$ and $R$ calculated by
  an SVD decomposition of a tensor network (c).  (e) Insertion of
  projection operators $P$ for a combination of $R$ and $L$. (f) A new
  renormalized tensor network. A new tensor $T'$ is the result of a
  tensor contraction of $R$ and $L$ and two $P$s in (e).}

We test our HOTRG algorithm based on EB operators in the calculation
of a partition function of the two-dimensional classical Ising
model. Tensor network representation of the partition function of the
two-dimensional Ising model is shown in \figref{HOTRG}(a). There are
two directions in a grid-type tensor network in \figref{HOTRG}(a). To
erase all loop entanglements in a renormalized scale, we apply the new
HOTRG procedure shown in \figref{BHOTRG} to two tensor $T$s linked
horizontally.  After that, we apply the conventional HOTRG procedure
to two tensor $T'$s linked vertically, because all loop entanglements
are already removed. The definition of a renormalization step in the
following is the pair of a new and a conventional HOTRG procedure for
horizontally and vertically linked tensors. We initially prepare a
tensor $T$ for $2 \times 2$ sites of the two-dimensional Ising
model. We set a limit of the bond dimension $D$ of tensor $T$'s
indexes. The limits of bond dimensions of a link $j$ and $k$ of an EB
operator $B$ in \figref{BHOTRG}(a) are $\sqrt{D}$ and $D$,
respectively. To solve the optimization problem stably, we initially
start the bond dimension of the link $j$ from 1, and we gradually
increase it to $\sqrt{D}$. For each bond dimension of the link $j$, we
also gradually increase a bond dimension of a link $a$ of a squeezing
operator $w$ in \figref{BHOTRG}(b) from 1. The limit is an effective
bond dimension (see the detail in Appendix \ref{ap:iteration}) of an
original link of a tensor $T$. If a loop entanglement flow exists, the
necessary bond dimension is less than the limit. In the increasing
step of the bond dimension of the link $j$ of $B$, we extend it as
like a squeezing operator $w$ in Appendix \ref{ap:iteration}. We
notice that the order of a computational cost to solve the
optimization problem does not change that of the total computational cost of a HOTRG
algorithm. The former is $O(D^6)$, and the latter is $O(D^7)$ (see the
detail of the computational complexity of the new HOTRG algorithm in Appendix
\ref{ap:complexity}).

\deffig{ISING_FREE}{0.48}{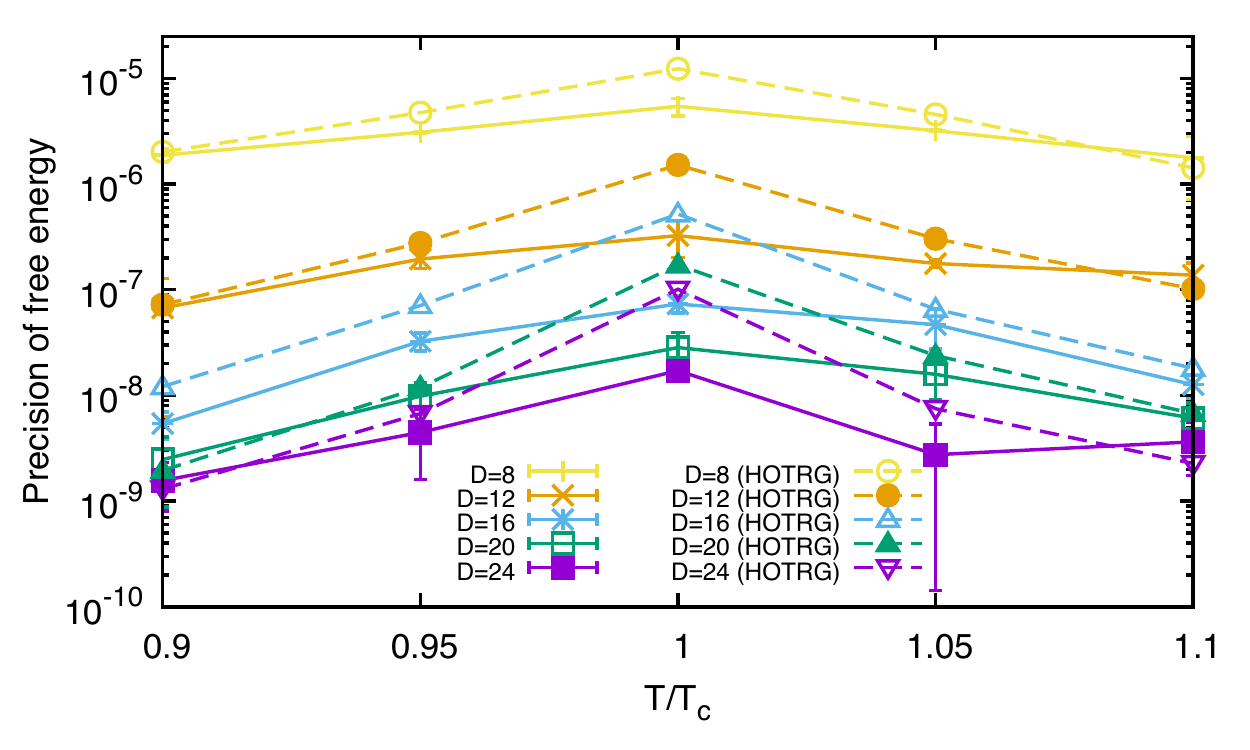}{Precision of
  free energy for the two-dimensional Ising model calculated by HOTRG
  algorithms with and without EB operation. A horizontal axis is a
  ratio of a temperature and a critical temperature $T_c$. $D$ is the
  limit of a bond dimension of tensor index in
  \figref{HOTRG}(a). Results of HOTRG calculations with and without EB
  operation are joined by solid and dashed lines, respectively.}
\deffig{ISING_BRANCHING}{0.48}{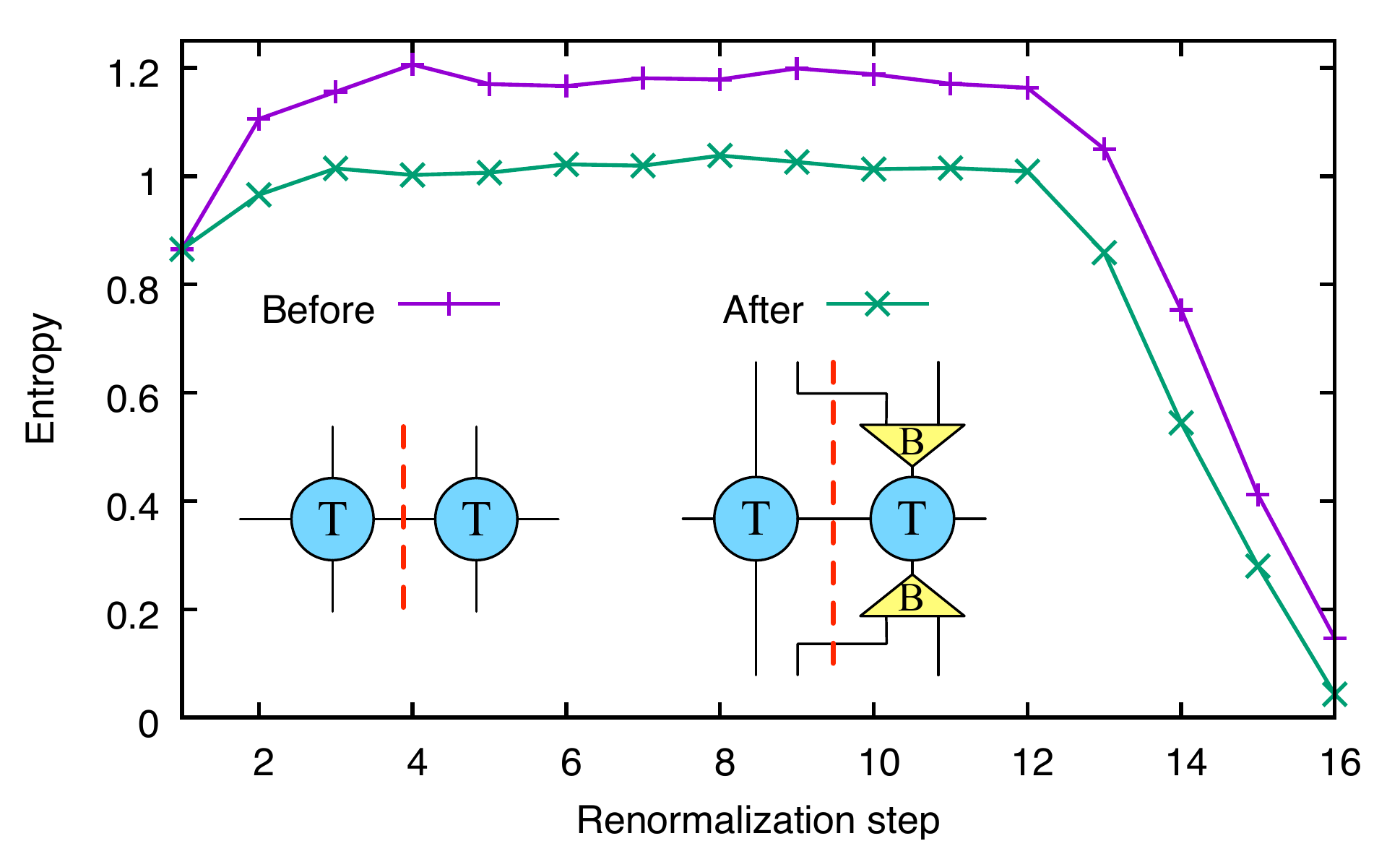}{Entropy
  of a composite tensor in a new HOTRG algorithm before and after EB
  operation at the critical point. The composite tensor is defined in
  the inset. A dotted line in the inset denotes a separation line
  between left and right parts of a composite tensor. Here, the limit
  of a bond dimension is $24$.}
\deffig{ISING_ENTROPY}{0.48}{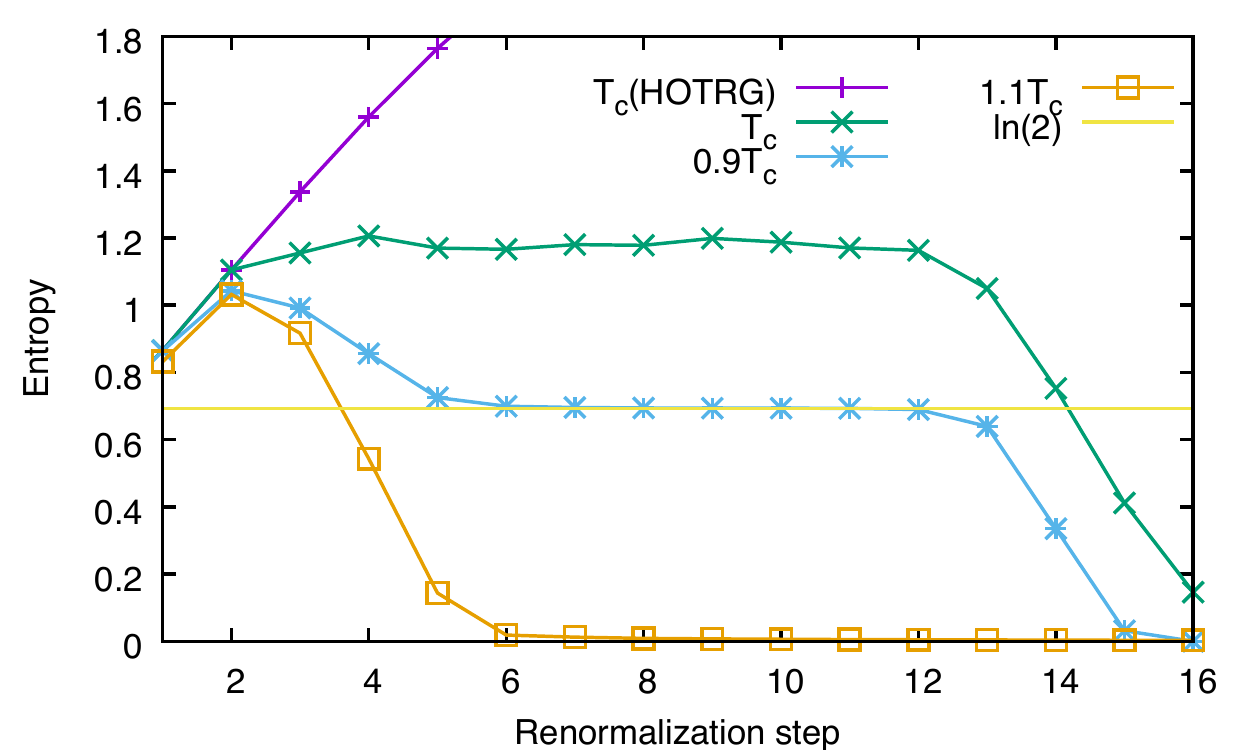}{Entropy of a
  composite tensor based on nearest neighbor tensors at three
  temperatures. $T_c$ is a critical temperature. Here, the limit of a
  bond dimension is $24$ in all cases.}
\figref{ISING_FREE} shows the precision of free energy calculated by
new and conventional HOTRG algorithms. Symbols joined by solid and
dashed lines denote the relative precision of free energy by HOTRG
algorithms with and without an EB operation, respectively. The
precision of the new HOTRG algorithm with the EB operation is better
than that of the original HOTRG algorithm at all temperatures. In
particular, the improvement is enhanced at the critical point
\footnote{At much far from a critical point, the precision of a new HOTRG algorithm with EB operators seems to be worse than that of the original HOTRG algorithm. At those points, there is not enough entanglement removed by an EB operator. Then, since the optimization problem is not well defined (see the inset in \figref{EB_EFF} of Appendix \ref{ap:iteration}), the obtained operator is not a proper EB. Therefore, an application of such operator may obstruct the coarse-graining procedure in a HOTRG algorithm.}
The reason is that the original HOTRG algorithm cannot erase entanglements
in a renormalized scale. To see the effect of an EB operator $B$, we
check an entanglement between two tensors. In the following, we define
an entropy of a normalized singular value distribution of a tensor as
$(-\textrm{Tr} \tilde{\Lambda} \log \tilde{\Lambda})$, where
$\tilde\Lambda = \Lambda/\textrm{Tr}\Lambda$. Here, $\Lambda$ is a
diagonal matrix of singular values for a matrix $M$ which is a matrix
representation of a tensor. Row and column indexes of $M$ denote left
and right parts of tensor indexes. The entropy of a composite tensor
defined by a tensor network is an estimator of an entanglement flow
through a link which connects two parts of a tensor
network. \figref{ISING_BRANCHING} shows the entropy of a composite
tensor in the new HOTRG algorithm before and after applying EB
operators at the critical point. Dashed lines in the inset of
\figref{ISING_BRANCHING} are cuts to define a decomposition into left
and right parts of a local tensor network before and after an EB
operation. The entropy after EB operations is reduced from the
original one. The entropy in \figref{ISING_BRANCHING} is reduced by
applying EB operators. The EB operator splits a target entanglement
flow correctly. Because of a decrease in an entanglement, we can
regard this procedure as a disentangling
operation. \figref{ISING_ENTROPY} shows the entropy of a composite
tensor based on nearest neighbor tensors (see the left tensor network
in the inset of \figref{ISING_BRANCHING}) at three temperatures. At
the critical point, the entropy does not increase. However, that of
the original HOTRG algorithm increases with the number of
renormalization steps as like that of the TRG algorithm. The behavior
of the new HOTRG algorithm is expected because we erase entanglements
in a renormalized scale for each renormalization step. In disordered
and ordered phases, the entropy converges to 0 and $\ln(2)$,
respectively. These values are consistent with fixed point tensors in
a disorder phase and an order phase.  All behaviors are similar to
that of TNR algorithm. From these results, we can confirm that the new
HOTRG algorithm using an EB operator catches a proper renormalization
flow in a tensor network space.

\deffig{BTN}{0.4}{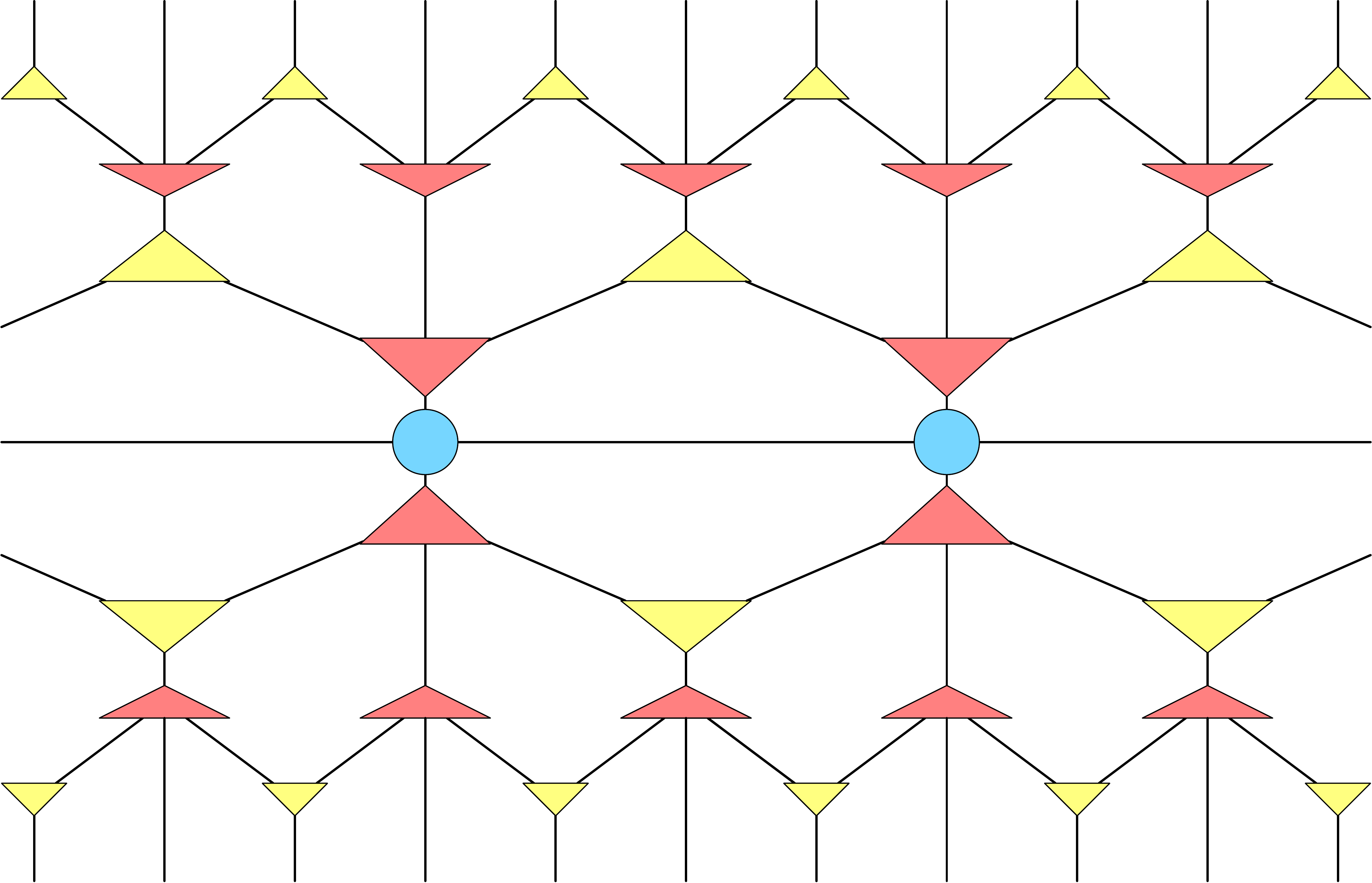}{A tensor network derived from a tensor
  network representation of a density operator by the HOTRG procedure
  with EB operation. A triangle with three links represents an EB
  operator. A triangle with four links represents a projection
  operator.  A circle with four links represents a coarse-grained
  tensor.}
Evenbly and Vidal discussed the derivation of MERA from a density
operator by using a TNR procedure\cite{Evenbly:2015ey}. The tensor
network representation of a density operator of a one-dimensional
quantum system is a grid-type tensor network shown in
\figref{HOTRG}(a). Also, there are two open boundaries along the
real-space direction. If we repeat a TNR procedure to the grid-type
tensor network with two open boundaries, we finally obtain the product
of two MERAs. Thus, we can derive MERA from a tensor network
representation of a density operator by TNR.  If we repeat a new HOTRG
procedure using an EB operator to a grid-type tensor network of a
density operator, we obtain a tensor network shown in
\figref{BTN}. Although a single link is split, the structure is
similar to that of MERA. This new tensor network state also holds the
log correction of the area law of entanglement entropy at a critical
point of a quantum chain as like MERA.

\subsection{Many-body decomposition}
\label{sec:MBD}
The conventional tensor decomposition is based on the matrix
decomposition. It transforms a tensor to a two-body tensor
network. For example, from (a) to (b) in \figref{TN}. Thus, there is a
limit of a transformation of tensor network topology.

\deffig{MBD}{0.48}{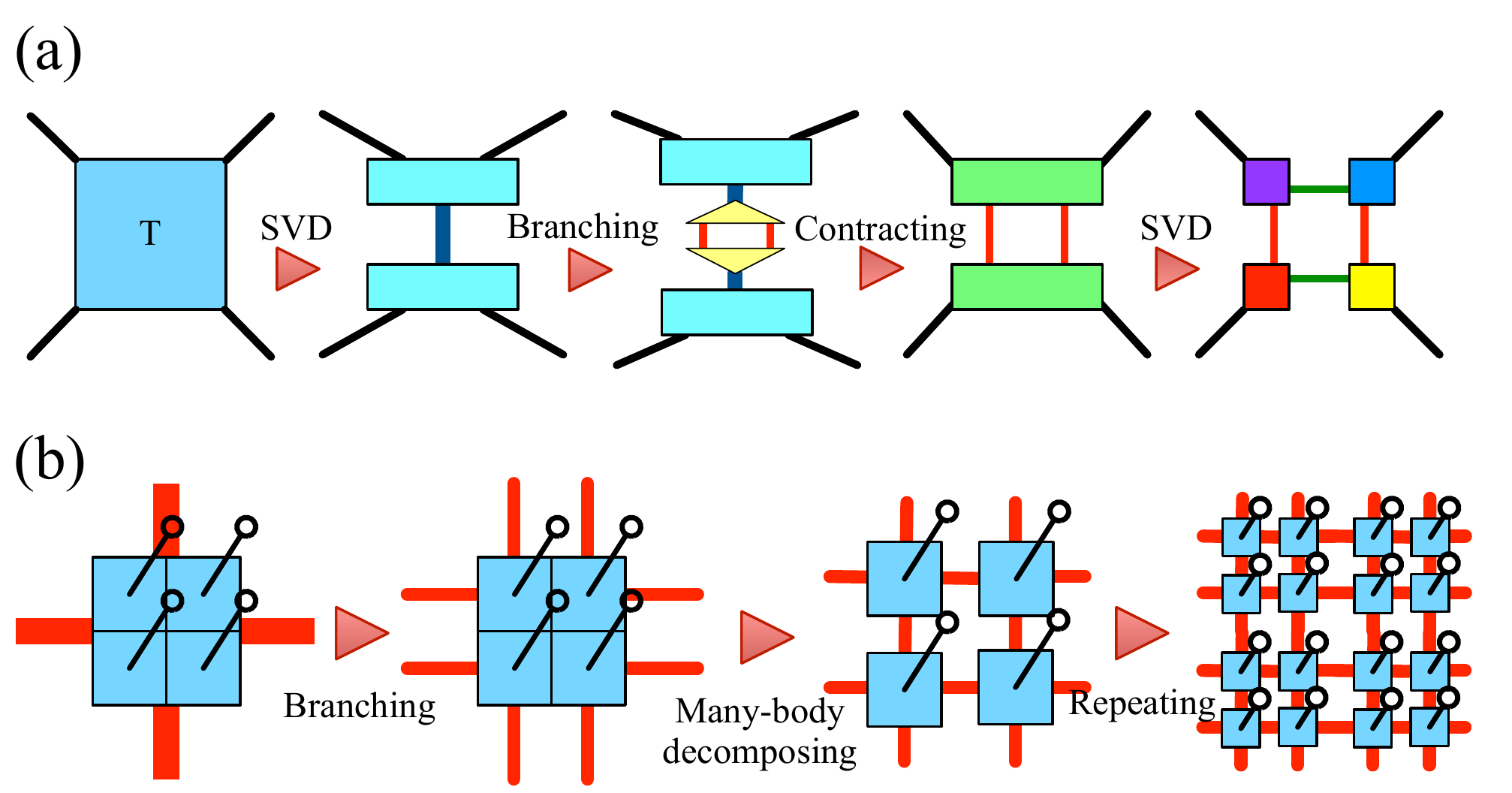}{(a) Many-body decomposition by EB
  operators. (b) Derivation of PEPS from a wave function.}

If we use an EB operator, we can transform a tensor to a many-body
tensor network as (c) in \figref{TN}. \figref{MBD}(a) shows a
procedure of a many-body decomposition.  At first, by using an SVD, a
tensor $T$ is decomposed into upper and lower tensors. It is a
conventional two-body decomposition. If we insert a pair of EB
operators on a contraction link between upper and lower tensors, we
can split a composite entanglement flow in the link into left- and
right-part entanglements. Contracting upper and lower tensors with EB
operators, we get new upper and lower tensors with new left and right
indexes. Decomposing new upper and lower tensors into sub-left and
sub-right tensors by using an SVD, we finally obtain a four-body
tensor network. This procedure defines a four-body decomposition with
a loop from a tensor $T$. It keeps a minimum entangled state on a loop
because an isolated loop entanglement does not exist in an initial
tensor. This procedure can be generalized for a many-body
decomposition. We notice that this procedure is an approximate
decomposition. We need to control the precision in the steps of
SVD. Under a given precision, a necessary bond dimension of a new link
depends on the strength of an entanglement flow.

The many-body decomposition may have interesting applications because
it gives us a new freedom to transform a topology of a tensor
network. The first application of a many-body decomposition is a
perfect disentangling for a loop entanglement structure. The
disentangling is an important idea for tensor network algorithms. For
example, a disentangler tensor in MERA is a key role in expressing a
critical quantum many-body state. The another example is a
disentangler tensor in TNR. It is crucial to reach a proper
fixed-point of critical phenomena by erasing a loop entanglement
structure. One way of a perfect disentangling for a loop entanglement
structure is a tensor contraction to erase a loop structure. We can
perfectly remove a loop entanglement structure in a four-body tensor
network of \figref{TN}(c) by tensor contractions to
\figref{TN}(a). There is no loop entanglement in a tensor of
\figref{TN}(a). To inverse this deformation, by using a four-body
decomposition shown in \figref{MBD}(a), we can again get a four-body
tensor network without a loop entanglement structure.  The second
conceptual application of a many-body decomposition is a systematic
derivation of PEPS from a wave function. \figref{MBD}(b) shows a
transformation of a tensor with four physical indexes which are
represented by links terminated by open circles. We first apply EB
operators to four unphysical indexes. EB operators split a composite
entanglement flow from two nearest neighbor physical indexes. If we
start from a wave function, we can skip this step, because there is no
unphysical index. Because we recursively apply this step to a part of
a derived PEPS in the following, we introduce this step.  Secondly, we
apply a variant of a four-body decomposition shown in
\figref{MBD}(a). Finally, we get a PEPS which consists of $2 \times 2$
blocks. If a physical index is composite in a block, we recursively
repeat this procedure. The many-body decomposition is approximate with
precision. Under a fixed precision, a bond dimension of a derived PEPS
depends on the strength of entanglements in a quantum state.  If a
quantum state satisfies the area law of entanglement entropy, we
intuitively expect that this derivation succeeds by a finite bond
dimension with accuracy. In fact, since the computational complexity
is huge, this derivation of a PEPS is conceptual. However, this
procedure shows that a metric of a tensor network state to describe a
quantum state can be related to entanglements in a quantum state.

\section{Conclusion and discussion}
\label{sec:SUM}
A tensor network and a tensor network algorithm grow new promising
theoretical tools to study various problems for many-body systems. To
add a new freedom for a tensor network algorithm, we proposed an EB
operation defined by an EB operator. It splits a composite
entanglement flow on a tensor index. We can set up an optimization
problem for splitting an entanglement flow by using a squeezing
operator. The optimization problem can be solved iteratively.

We introduced two applications of an EB operation.  The first one is
an improvement of HOTRG algorithm to catch a proper renormalization
flow in a tensor network space. The numerical results for the
two-dimensional Ising model show expected properties in a precision of
a free energy calculation and a local entanglement between two
coarse-grained tensors. We also derived a new tensor network state
from applying our improved HOTRG procedure to a grid-type tensor
network of a density operator. The second application is a many-body
decomposition of a tensor. Using it, we can change a topology of a
tensor network directly. We can apply it to a perfect disentangling
and a systematic derivation of a PEPS from a wave function.

The purpose of an EB operation is to split a composite entanglement
flow on a link in a tensor network. We can use it for a disentangling
in a part of a tensor network as shown in the case of the improved
HOTRG algorithm. Thus, the EB operator may be regarded as a
disentangler in MERA and TNR. However, the purpose of a disentangler
is different from that of an EB operator. It is a disentangling
between two local degrees of freedom in a tensor network. In fact, a
disentangler tensor does not consist only of a role of EB
operation. For example, a disentangler tensor in TNR may contain both
roles of projection and an EB. The disentangler is an important
concept for tensor networks. The EB is a new basic operation which can
be applied to the implementation of the disentangler. It may have
other applications as a many-body decomposition.

From a practical point of view, the computational cost to optimize an
EB operator is an issue. In particular, the number of iterations in
the iteration method (see Appendix \ref{ap:iteration}) is a
problem. In fact, in the case of the two-dimensional Ising model, we
need more than 1000 iterations to solve the optimization problem of an
EB operator. In the case of improved TRG algorithms, the computational
cost of a loop optimization technique \cite{Yang:2017hj} and a Gilt
technique \cite{Hauru:2018ij} is much less than that of
TNR\cite{Evenbly:2015cs, Evenbly:2017en}. Thus, it extends the
application scope of an improved TRG algorithm in a real study. For an
optimization of an EB operator, we also need to reduce a total
computational cost. Although we start from randomized initial tensors,
there may be good initial tensors. To avoid a local solution, we
extend tensor size gradually. There may be a good iteration
strategy. The improvement of solving the optimization problem of an EB
operator remains much for future research.

Since the improved algorithms based on TRG as like TNR mix space and
(imaginary-)time directions, they cannot be directly applied to
anisotropic cases. However, the improved HOTRG algorithm with an EB
operation can be applied to such problem, because it only does a
coarse-graining of tensors along a chosen direction. Based on the same
property, the HOTRG algorithm was extended to a three-dimensional
grid-type tensor network\cite{Xie:2012iy}. The extension of our
approach to a three-dimensional case is also interesting.

\begin{acknowledgments} The author appreciates conversations with
N.~Kawashima, S.~Morita, and T.~Suzuki, and, very specially, with
T.~Okubo, whose comment for the TNR algorithm was crucial to
considering the optimization problem for an EB operator. This work was
supported by JSPS KAKENHI Grants No. 26400392 and No. 17K05576, and by
MEXT as ”Exploratory Challenge on Post-K computer” (Frontiers of
Basic Science: Challenging the Limits) and "Priority Issue on Post-K
computer” (Creation of New Functional Devices and High-Performance
Materials to Support Next-Generation Industries).
\end{acknowledgments}

\newpage
\appendix
\section{Iteration method to optimize an entanglement branching
operator}
\label{ap:iteration}
To optimize an EB operator in \figref{BR}(a), we need to minimize a
distance between two tensor networks of \figref{BR}(a) and (c). An EB
operator $B$ and squeezing operators $w$ and $v$ are isometric. Thus,
the minimization between two tensor networks of \figref{BR}(a) and (c)
is a maximization of a norm of a four-body tensor network by $B$, $T$,
$w$ and $v$ as shown in \figref{IM}.
\deffig{IM}{0.48}{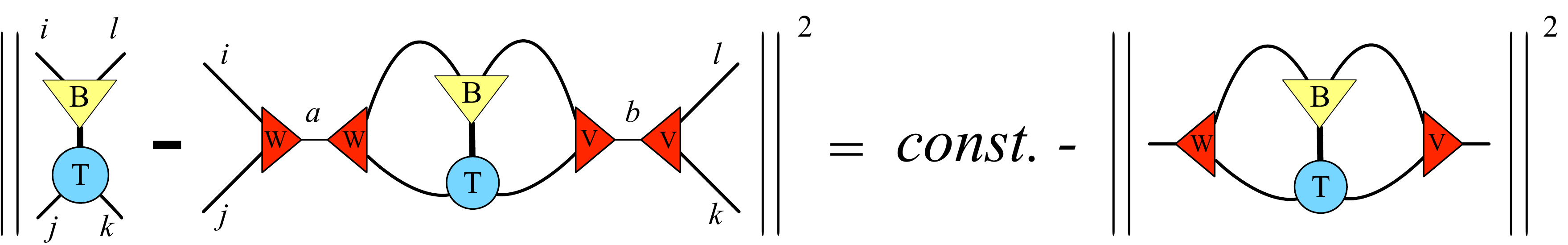}{A squared distance between two tensor
  networks to optimize an EB operator.}

We can solve the maximization problem by iteration updates of $B$,
$w$, and $v$. If we fix tensors except for a tensor $w$ or $v$, the
maximization problem for $w$ or $v$ can be written as a
diagonalization problem. If we fix tensors except for a tensor $B$, we
can solve the maximization problem for $B$ by using an SVD
optimization technique for MERA\cite{Evenbly:2009bk}.

However, there may be many local solutions in the total maximization
problem. To avoid a local solution, we use a strategy to extend a
solution of $w$ and $v$ gradually. The procedure is written as
follows:
\begin{enumerate}
\item Initialize $B$ randomly.
\item Set the values of bond dimensions of links $a$ and $b$ 1, and
initialize $w$ and $v$ randomly.
\item Iteratively update $B$, $w$, and $v$ to minimize the squared
distance between \figref{BR}(a) and (c). Because they are isometries,
the local optimization problem for a tensor $B$ can be solved by the
singular value decomposition method as the optimization of isometries
in MERA\cite{Evenbly:2009bk}, and it for a tensor $w$ or $v$ can be
solved by a diagonalization of an environment of a target
tensor. Here, we define an environment as a composite tensor of which
a tensor contraction with target tensors is a maximized squared
norm. \label{opt}
\item Increase bond dimensions of links $a$ and $b$ (extend bond
dimensions of $w$ and $v$). New elements of $w$ and $v$ are
initialized as zero, but other elements are unchanged.
Alternatively, we can increase a bond dimension in a diagonalization of an environment of a target tensor, $w$, and $v$, respectively.
\item Go back to Step. \ref{opt}, until bond dimensions of links $a$
and $b$ reach a limit of them.
\end{enumerate}

We can estimate the limit of bond dimensions of a link $a$ and $b$ by
an entropy of a tensor $T$ between an index of a target link and a
composite index of other links.

\deffig{EB_EFF}{0.48}{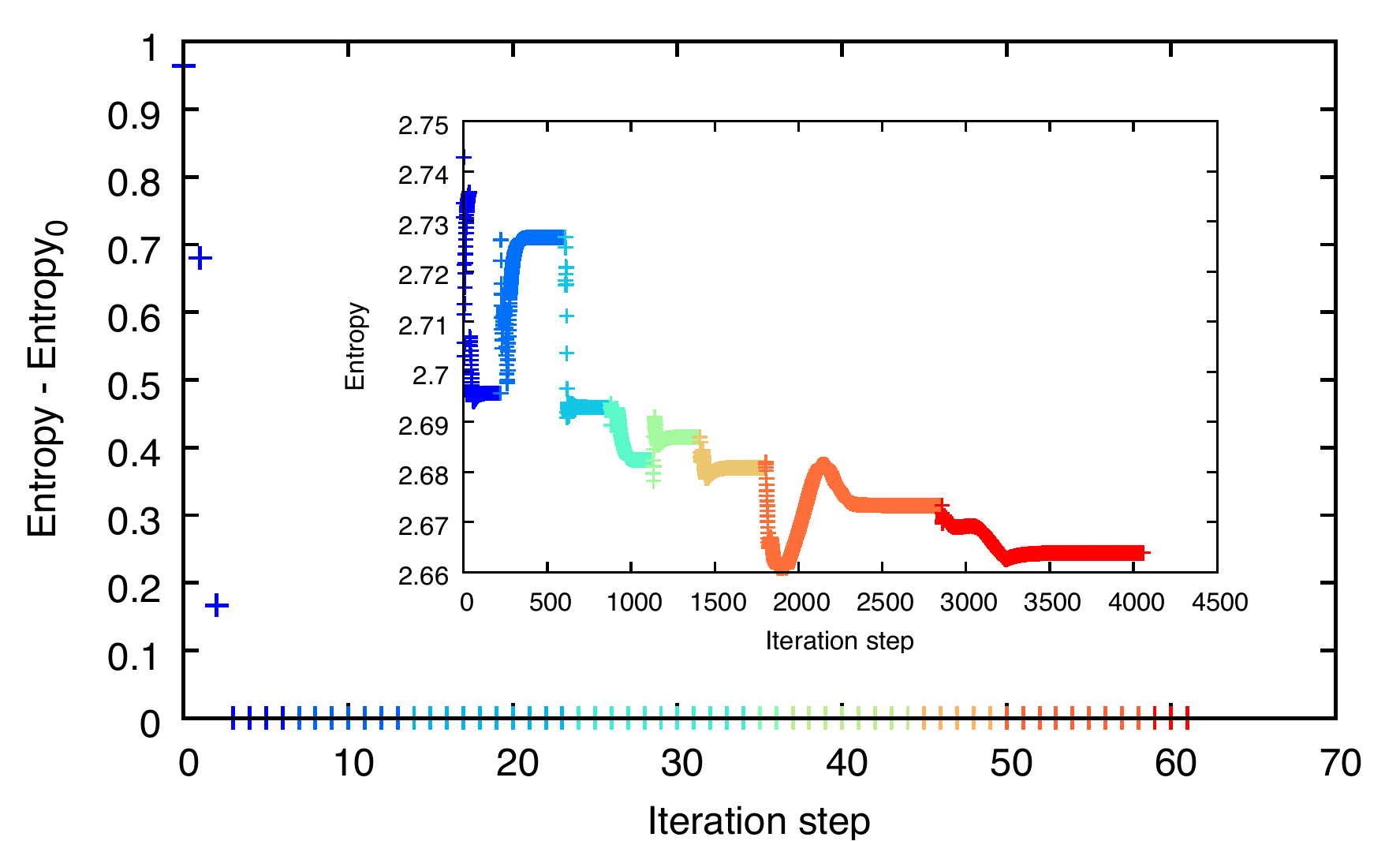}{Entropy profile of a
  composite tensor of $B$ and $T$ in an optimization process of EB
  operator in \figref{IM}. The main panel shows a result of a CDL
  tensor with a random unitary on a target index. The bond dimension
  is $D=3\times 3$. Entropy${}_{0}$ is $\log(3)$. The inset shows that
  of a random tensor for the same bond dimension. All tensors, $B$,
  $w$, and $v$, are randomly initialized. We start an initial bond
  dimension of links $a$ and $b$ from one. A color of a symbol denotes
  a bond dimension of links $a$ and $b$.}

\figref{EB_EFF} shows entropy profiles in the above optimization
process of EB operator in \figref{IM}. The main panel shows a result
of a CDL tensor with a random unitary on a target link of EB operator as follows:
\begin{equation}
  \label{eq:CDL}
T_{i_1i_2, j_1j_2, k_1k_2} = U_{i_1i_2, i'_1i'_2}\delta_{i'_1, j_2}\delta_{i'_2, k_2}\delta_{j_1, k_1},
\end{equation}
where $U$ is a random unitary and the composite index $(i_1,i_2)$ is a
target of EB operator. Thus, the ideal EB operator is
$U^{\dagger}$. When the bond dimension of a sub index is $\sqrt{D}$,
the entropy of a composite tensor of the ideal $B$ and $T$ is
$\log(D)/2(=\mbox{Entropy}_0)$ when it is decomposed into a left index
group $i$ and $j$ and a right index group $l$ and $k$. As shown in the
main panel in \figref{EB_EFF}, the entropy rapidly converges to the
ideal value. A color of a symbol denotes a bond dimension of links $a$
and $b$. Although we cannot expect an proper EB operator for a general
random tensor $T$, the optimization method struggles to find a better
EB operator. But, even for a random tensor, the optimization process
decreases the entropy of a composite tensor as shown in the inset of
\figref{EB_EFF}. Therefore, the proposed optimization method of EB
operator is efficient.

\section{Computational complexity of a new HOTRG algorithm with entanglement
branching operators}
\label{ap:complexity}
The procedure of the new HOTRG algorithm with EB operators consists of
three parts: (i) an optimization of an EB operator, (ii) a calculation
of new tensor $L$ and $R$, and (iii) a calculation of a coarse-grained
tensor from $L$ and $R$. For simplicity, we suppose that the bond
dimension of tensor indexes except for a link $b$ in
\figref{CBHOTRG}(a) is $D$ and the bond dimension of the link $b$ is
$\sqrt{D}$.

\deffig{CBHOTRG}{0.48}{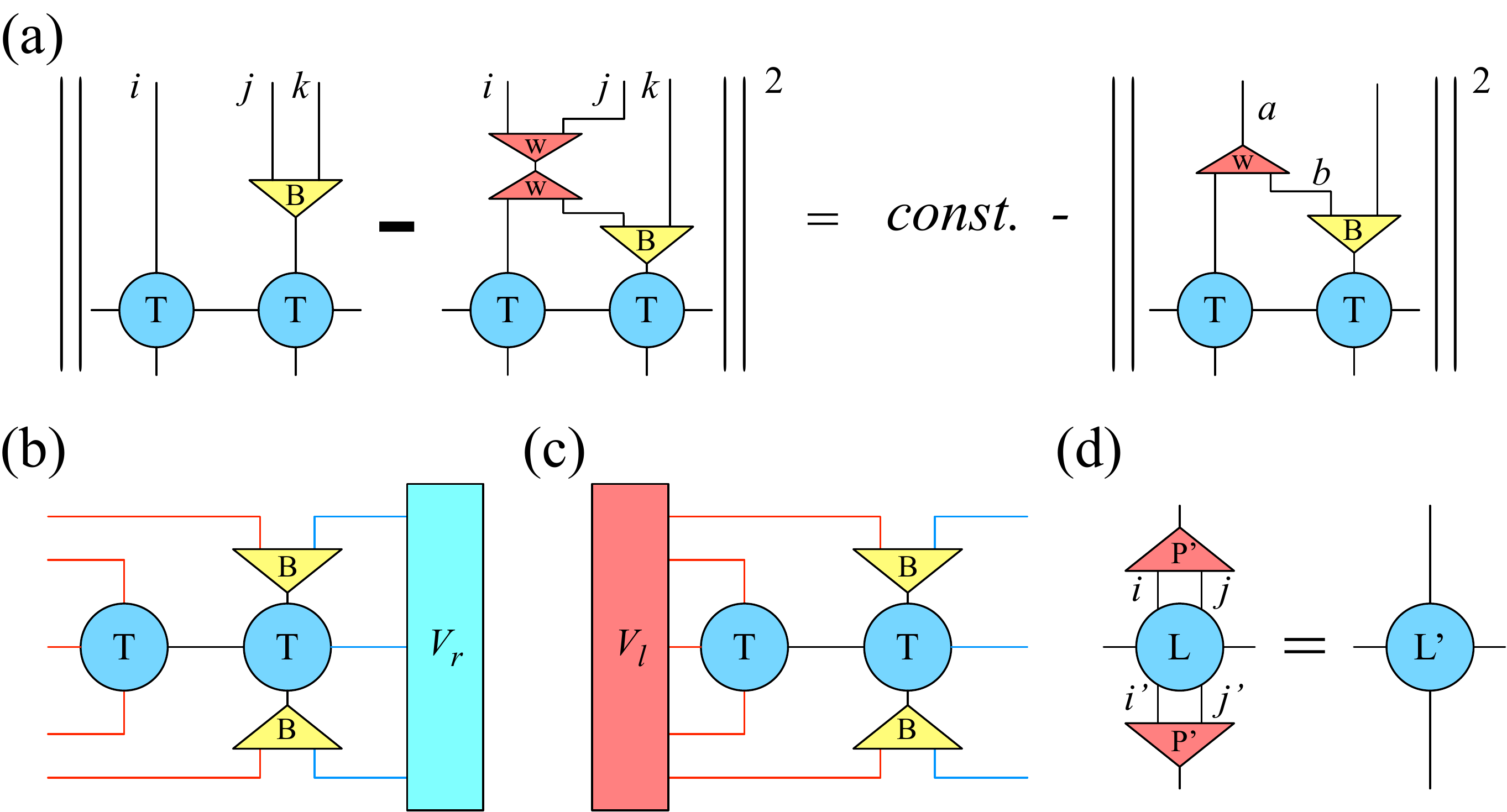}{The calculation in a new HOTRG
  algorithm with EB operators.  (a) A squared distance between two
  tensor networks to optimize an EB operator.  (b) A right
  matrix-vector multiplication.  (c) A left matrix-vector
  multiplication.  (d) A coarse-grained tensor $L'$.  }

The first part (i) is to solve a maximization of a squared norm of a
tensor network by $B$, $T$, $w$ as shown in the right squared norm in
\figref{CBHOTRG}(a). Since $B$ and $w$ are isometries, we can use an
iteration method based on an SVD optimization technique as like
MERA\cite{Evenbly:2009bk}. Thus, the computational complexity of the first part (i)
is governed by the calculation of environments for an SVD optimization
technique. The environment is a composite tensor defined by a tensor
network which is a representation of the squared norm in the right
part of \figref{CBHOTRG}(a) except for a target tensor. The computational complexity
of the calculation of an environment is $O(D^6)$. Also, the total
computational time is proportional to the number of iterations to
update tensors in the iteration method.  As explained in Appendix
\ref{ap:iteration} and \secref{BHOTRG}, we gradually increase the bond
dimension of the link $a$ and $b$ in \figref{CBHOTRG}(a).

In the second part (ii), we use a partial SVD algorithm for the tensor
network in \figref{BHOTRG}(c) to decompose it into $L$ and $R$ in
\figref{BHOTRG}(d). We need to calculate a right and left
matrix-vector multiplication for the partial SVD algorithm.  They are
shown in \figref{CBHOTRG}(b) and (c). Here, $V_r$ and $V_l$ are right
and left vectors, respectively. The computational complexity of their matrix-vector
multiplications is $O(D^5)$. Thus, the total computational complexity of a partial
SVD algorithm is $O(D^6)$.

In the third part (iii), we introduce an intermediate tensor $L'$
which is applied to a projection operator for upward and downward
indexes, $i, j, i'$ and $j'$ of $L$. The bond dimension of upward and
downward indexes of $L'$ is $D$. The computational complexity of the calculation of
$L'$ is $O(D^6)$. Also, the computational complexity of the calculation of the
coarse-grained tensor $T'$ in \figref{BHOTRG}(e) from $L'$ and $R$ is
$O(D^7)$.

The maximum computational complexity is the third part (iii). Therefore, the total
computational complexity of the new HOTRG algorithm with EB operators is $O(D^7)$.

\section{Critical fixed-point tensor of a new HOTRG algorithm with
  entanglement branching operators}
\label{sec:exponents}
When we apply a new HOTRG procedure with EB operators to a
renormalized tensor at a critical point, it quickly converges to a
critical fixed-point tensor as shown in \figref{ISING_ENTROPY}. There
are several methods which derive a universal data from a critical
fixed-point tensor. In particular, for a two-dimensional critical
system, Gu and Wen\cite{Gu:2009} proposed a useful method based on a
conformal field theory. Then, the scaling dimension can be estimated
from eigenvalues of a transfer matrix constructed from a critical
fixed-point tensor as follows:
\begin{equation}
  \label{eq:gu_wen}
  \Delta_i = -\frac{1}{2\pi}\log(\lambda_i/\lambda_0),
\end{equation}
where $\lambda_i$ is the $i$-th eigenvalue of a transfer matrix
defined by a renormalized tensor and $\lambda_0$ is the largest
eigenvalue. \figref{SCALING_DIMENSION} shows scaling dimensions by
\eqref{gu_wen} at a renormalization step for the original HOTRG
algorithm and the new one. We construct the transfer matrix from two
columns of tensors ($L=2$ transfer matrix in
ref. \cite{Yang:2017hj}). The bond dimension $D$ is 24 in both
cases. The high-level scaling dimensions of the original HOTRG
algorithm start to merge with the low-level scaling dimensions after
three renormalization steps. However, those of the new HOTRG algorithm
with EB operators keep up to ten renormalization steps with $2^{22}$
spins. Therefore, EB operators improve a critical property of a
renormalized tensor.

\deffig{SCALING_DIMENSION}{0.48}{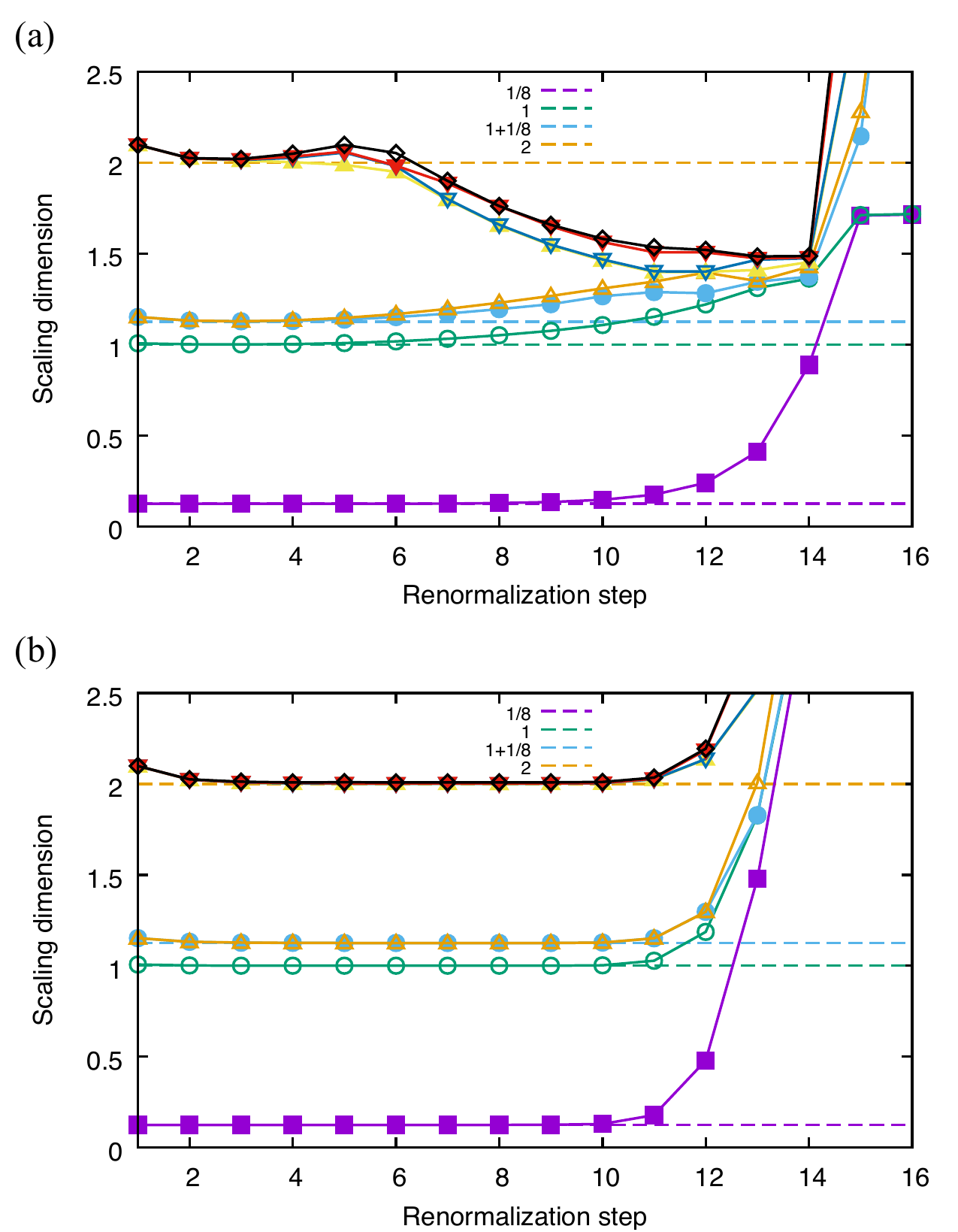}{Comparison of
  scaling dimensions for the original HOTRG and the new HOTRG
  algorithm with EB operators.  (a) Scaling dimensions of the original
  HOTRG algorithm at a renormalization step. (b) Scaling dimensions of
  the new HOTRG algorithm with EB operators at a renormalization
  step. A transfer matrix is constructed from two columns of tensors
  ($2 \times 2$ tensors). The bond dimension $D$ is 24 in both cases.
  Dotted lines denote exact values of scaling dimensions of the
  two-dimensional Ising model. }
\begin{table}[h]
  \centering
  \begin{tabular}[t]{c|c|c}
    & exact & HOTRG with EB op.\\
    \hline
    $c$ & 0.5 & 0.49996(2)\\
    $\sigma$ & 0.125 & 0.12515(3)\\
    $\epsilon$ & 1 & 1.0002(1) \\
    & 1.125 & 1.1250(1)\\
    & 1.125 & 1.1252(1)\\
    & 2 & 2.0009(2)\\
    & 2 & 2.0013(2)\\
    & 2 & 2.0029(4)\\
    & 2 & 2.008(1)
  \end{tabular}
  \caption{Exact values and numerical estimation of scaling dimensions
    and a central charge from a renormalized tensor by the new HOTRG
    algorithm with EB operators. A transfer matrix is constructed from
    two columns of tensors ($2 \times 2$ tensors). The bond dimension
    $D$ is $24$. The last digit with a bracket means a confidential
    interval estimated from seven ($2^{16}$ spins) to nine ($2^{20}$
    spins) renormalization steps .}
  \label{tab:scaling_dimensions}
\end{table}
Table \ref{tab:scaling_dimensions} shows the estimated values of
scaling dimensions and a central charge of the new HOTRG algorithm
with EB operators. The accuracy is comparable with the other
entanglement-filtered tensor network algorithm (for example, see
Tables in Refs. \cite{Evenbly:2015cs, Yang:2017hj, Hauru:2018ij,
  Bal:2017dp}).

\bibliography{tensor}
\end{document}